\documentclass[preprint,12pt]{elsarticle}

\usepackage{graphicx}

\usepackage{amssymb}
\usepackage{amsmath}

\journal{Nucl. Instr. Meth. A}

\begin{document}

\begin{frontmatter}



\title{Point spread function due to multiple scattering of light in the atmosphere}


\author{J. P\c ekala \corref{cor1}}
\ead{Jan.Pekala@ifj.edu.pl}

\author{H. Wilczy\'nski \corref{cor2}}
\ead{Henryk.Wilczynski@ifj.edu.pl}

\cortext[cor1]{Corresponding author. Tel.: +48 12 662 8341; fax: +48 12 662 8012.}

\address{Institute of Nuclear Physics PAN, Radzikowskiego 152, 31-342 Krak\'ow, Poland}

\begin{abstract}
The atmospheric scattering of light has a significant influence on results of optical observations of air showers. It causes attenuation of direct light from the shower, but also contributes a delayed signal to the observed light. The scattering of light therefore should be accounted for, both in simulations of air shower detection and reconstruction of observed events. In this work a Monte Carlo simulation of multiple scattering of light has been used to determine the contribution of the scattered light in observations of a point source of light. Results of the simulations and a parameterization of the angular distribution of the scattered light contribution to the observed signal (the point spread function) are presented.
\end{abstract}


\end{frontmatter}



\section{Introduction}

The observation of light produced by an extensive air shower is one of the established methods of detection of ultra-high energy cosmic rays. As the cascade of energetic charged particles of an air shower, initiated by a highly energetic primary cosmic ray particle, moves through the air, it produces a large number of fluorescence and Cherenkov photons \cite{auger10, hires02, ta12}. The intensity of the fluorescence light \cite{airfly07, arqueros09} is proportional to the size of the air shower at any point of its trajectory, therefore observations of light along the shower path enable determination of a profile of air shower development, which provides information about the properties of primary cosmic ray particles. If sufficiently large portion of air shower development is observed, it is possible to obtain by simple integration a very precise, model independent estimation of energy of the primary cosmic ray particle. Properties of air showers differ for different primary particles, therefore the observed profiles of air shower development, especially the atmospheric depth where the number of particles in the shower reaches its maximum, provide valuable information about cosmic ray composition.

While the direct fluorescence light can be regarded as a signal that can be easily interpreted, there are also other contributions to the observed light. The Cherenkov light is emitted mostly at small angles relative to the shower axis, and so only in small fraction of events, when the air shower direction points very close to the detector, it is observed directly in significant amounts. However, both the Cherenkov photons as well as the isotropically emitted fluorescence photons, regardless of their initial directions, can get scattered in air, and with some probability be directed towards the detector. This scattered light is delayed with respect to the direct light emitted simultaneously, and it may also come from different directions, according to the position where the photons last scattered. As the shower progresses, the scattered light emitted at different times and positions along the shower path overlay each other, forming a halo that follows the image of the air shower on the sky. Indeed, a faint signal is recorded from directions distant from the location of the light source \cite{icrc11, parrisius09, werner10}. Until now there is no conclusive explanation of this observed halo. Different effects are investigated as potential sources of this halo: the properties of the detector itself, and also the light scattering in different atmospheric conditions. In particular the aerosols of large sizes may be responsible for at least a part of this effect \cite{louedec11}.

Previous analyses \cite{roberts05,pekala09,giller12} focused on describing the contribution of the scattered light to the observed shower image integrated within some radius around the center of the shower image. In this case a light source moving with the speed of light through the atmosphere is considered. The longitudinal shower profile and different geometries of observation are considered, according to the conditions at which the real air showers are expected to be observed. This integral contribution needs only to be known in a small region of the sky around the position of the air shower, at the same time when the shower is observed. 

The aim of this work is to describe a more general case of a point-like, stationary light source. Assuming a finite emission time, the propagation of light in the atmosphere is simulated, giving information about the distribution of light that is observed by a distant detector at different moments of time. In simulations of air shower detection, one calculates the light emitted at different points along the shower path, and based on this the signal recorded by the detector is calculated. Previous parameterizations of the total (integrated) signal of scattered light that arrives to the detector, including contributions from different points of emission, can not be applied. In this case only parameterization of scattered light from a point source can be applied - combined with distribution of emission along the shower path it will enable more accurate simulations of air shower observations. Another application of such parameterization is in analysis of dedicated measurements of the detector response done with the help of airborne light sources. As mentioned above, such measurements are done \cite{parrisius09, werner10}, and a description of scattered light from a point source is necessary for their interpretation.

The point spread function, which we developed in this work, describes the distribution of the scattered photons on the sky as a function of time. This function describes the light as it arrives to the detector, without simulating the detector response. To be most useful, such a parameterization needs to cover not only the immediate vicinity of the light source location, but also more distant areas of the sky, because for a moving source, the most important area will shift correspondingly. The distributions of the scattered light must be described not only at a single time bin simultaneous with the observation of the shower at the same direction, but for a more extended period of time. This is necessary, because we need to know the different contributions (with different individual delays, and different times of emission) that coincide at the moment of observation. The parameterization developed in this work requires only a few basic parameters describing the conditions of observation, so it can be easily integrated into any computer program dedicated to studies of light sources in the atmosphere. Its application should help decrease the systematical uncertainty of the energy estimations in air shower experiments using fluorescence detectors.

While this work is dedicated to study the effects of scattering of light, it is worth noting that also absorption should be taken into account. This however would require detailed knowledge about aerosols at the sites of observation, since absorption depends on aerosol composition, as well as the light wavelength.

\section{Simulations}

In this work we used the ``Hybrid\_fadc'' program \cite{dawson96}, which was originally designed to simulate air showers. In this framework a Monte Carlo program has been previously developed for the analysis of multiple scattering of light and its effects on observations of air showers \cite{pekala09}. A new version of this program was prepared, which allows us to simulate a point source of light and the subsequent propagation of photons through the atmosphere, and finally to obtain the distribution of light arriving at a detector.

The location of the light source must be chosen, specified by distance from the location of the detector and altitude above ground. The amount of light emitted can also be specified, although in our analysis this value is not important since all scattered signals were considered as relative to the intensity of the light coming directly from the source. Also the duration of emission can be chosen: an instantaneous emission, or a continuous one for any time period. In all our analysis the emission was assumed to be isotropic, imitating fluorescence emission from air showers, but it would be also possible to start with another angular distribution of emission if required. The detector that records the light is located at the ground level, and the observed light is recorded in simulation as it arrives at a selected position - without effects of the observation process itself, which would be different for specific detectors. The ground level was set at an altitude of 1570 m above sea level, which is roughly the mean of the altitude range of fluorescence detectors of HiRes and Pierre Auger Observatory.

In the simulation, packets of photons are traced on their way through the atmosphere. For any direction the conditions along the path of photons (the density of scattering centers), and based on that distances at which scattering occurs are randomly chosen - one for Rayleigh scattering and one for scattering on aerosols only. From these two distances, the one which occurs closer to the starting point is taken as the one that really takes place - this way not only the position, but also the process responsible for the individual scattering is determined. The light may get scattered on air molecules, which is described by the Rayleigh scattering, or on aerosols, which can be approximated by different models of aerosols in the framework of Mie scattering. Once the random procedure determines the mechanism of scattering, it is possible, based on the angular distribution and distance to the detector, to calculate what fraction of the scattered light is observed. In order to trace those photons of the scattered packet that were not directed to the detector, it is assumed that they continue their flight together. For such a smaller packet the calculations are repeated several times to account for further scatterings. The program then records distributions on the sky of arriving light, separately for different time bins after the start of observations.

In the simulation it may happen that the point of scattering is randomly placed very close to the detector. In such a case, the geometrical factor (distance squared) causes that a very large fraction of the scattered photon packet is recorded from this single direction and stands out in the final results. To minimize this effect, all simulations were repeated ten times for identical geometries and atmospheric conditions, and the results were averaged.

In our analysis the source was assumed to emit the monochromatic light of 370 nm wavelength, which is in the near ultraviolet range used in fluorescence detectors of air showers. 100 ns was used as the length of time bins for collection of light at the detector - this is of the order of time resolution of fluorescence detectors of air showers \cite{auger10, hires02, ta12}. As we aim to model the conditions of observation of air showers, where the observed light intensity varies relatively slowly in this time scale, the light is assumed to be emitted with constant intensity. To enable easy integration of signals with different delays, the light is emitted also over the period of 100 nanoseconds.

The simulations were done using an atmosphere described by the US Standard Atmosphere Model \cite{usstd}. Aerosols were assumed to have a simple exponential vertical distribution, with height scale of 1.2 km. As it was shown \cite{pekala09}, different distributions of molecular atmosphere and aerosol scale heights have no significant effects on the results, when the difference in optical distance is taken into account. Various light attenuation lengths at ground level $\Lambda_{T}$ correspond to different concentrations of aerosols in air, from a dirty atmosphere in which the scattering is dominated by aerosols to almost purely molecular one.

The light scattering simulations were done for an extensive set of atmospheric conditions and geometries, namely for all combinations of:
\begin{itemize}
\item horizontal distance from the detector: 3, 7, 15, 25, 35 km;
\item altitude above ground: 1, 2, 3, 4, 5, 6, 7, 10 km;
\item total horizontal attenuation length $\Lambda_{T}$: 6.6, 10, 12.6, 17, 19.9 km (they correspond to $\Lambda_{Mie}$ 9.6, 19.2, 32, 96 and 480 km respectively for a constant $\Lambda_{Rayleigh}$ equal 20.7 km);
\item four different aerosol phase functions.
\end{itemize}

To investigate how aerosols of different sizes scatter the light, we have used different aerosol phase functions (i.e. angular distributions of scattering). And so we used the phase function that was used in our previous analyses \cite{pekala09,dawson96}, which is based on the desert aerosol phase function of \cite{longtin88} (Longtin phase function), and also three more described by a modified Henyey-Greenstein function \cite{henyey41,louedec12}

\begin{equation}
 P_a(\alpha; g,f) = \frac{1-g^2}{4\pi} \left[ \frac{1}{(1+g^2-2g \cos \alpha)^{3/2}} + f \frac{3 \cos^2 \alpha -1}{2(1+g^2)^{3/2}} \right].
\end{equation}

\begin{figure}[ht]
\begin{center}
\includegraphics[scale=0.9]{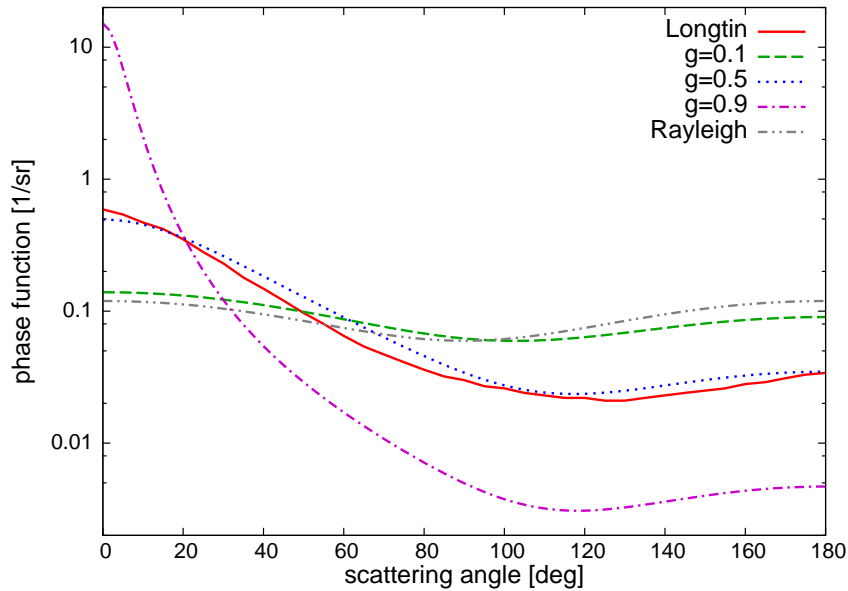}\\
\caption{\label {fig8} Aerosol and Rayleigh phase functions used in the simulations.}
\end{center}
\end{figure}

The parameter $g$ = $\langle \cos \alpha \rangle$ describes the asymmetry of the scattering distribution in the forward/backward direction and its value is determined by size of aerosols that are present in the air at the moment of observation. Larger values of parameter $g \gtrsim 0.8$ correspond to a mean aerosol size of several micrometers, while smaller values of $g$ describe scattering on smaller, submicron aerosols. It is worth noting that the relation between $g$ and mean aerosol size is wavelength dependent, which means that for observations in identical atmospheric conditions different values of the $g$ parameter apply to different light wavelengths \cite{louedec09, louedec12ao}. In this analysis we have repeated all simulations for aerosol phase functions corresponding to $g$ equal to 0.1, 0.5 and 0.9. The parameter $f$ describes the small backward scattering peak, not accounted for by the original Henyey-Greenstein function \cite{henyey41}. All simulations were done for $f$ = 0.4, which best describes the observed properties of scattering on aerosols (large $f$ may result in negative values of the phase function, but for the value used it is not the case). All phase functions used in simulations are shown in fig. \ref{fig8}.

\section{Results of simulations}

As a result of the simulation, we get a series of distributions of light on the sky, describing the image that can be observed by a detector, in different time bins of 100 ns length. The light was recorded for a total time of 5 $\mu s$, starting from the moment when the first direct photons arrive at the detector ($t$ = 0). Our previous work \cite{pekala09} showed that the image formed by the scattered light on the sky is with a good approximation radially symmetrical, with the center at the position of the light source. Therefore, instead of a two-dimensional image of the sky, we may consider the radial distribution, as a function of angle $\zeta$ from the location of the light source on the sky.

To allow an easy comparison, the scattered light contributions presented in this work are calculated relative to the direct signal received from the light source. This direct signal is calculated based on the distance from the detector, and accounts for attenuation of light due to scattering of light along its path. Since the simulated light source emits for a period of 100 ns, this means that only in the first time bin one could really compare the intensities of light. In later time bins no direct light is received, but for the sake of uniform normalization, the scattered light contribution in all time bins is calculated relative to the direct light in the first time bin.

\begin{figure}[ht]
\begin{center}
\includegraphics[scale=0.9]{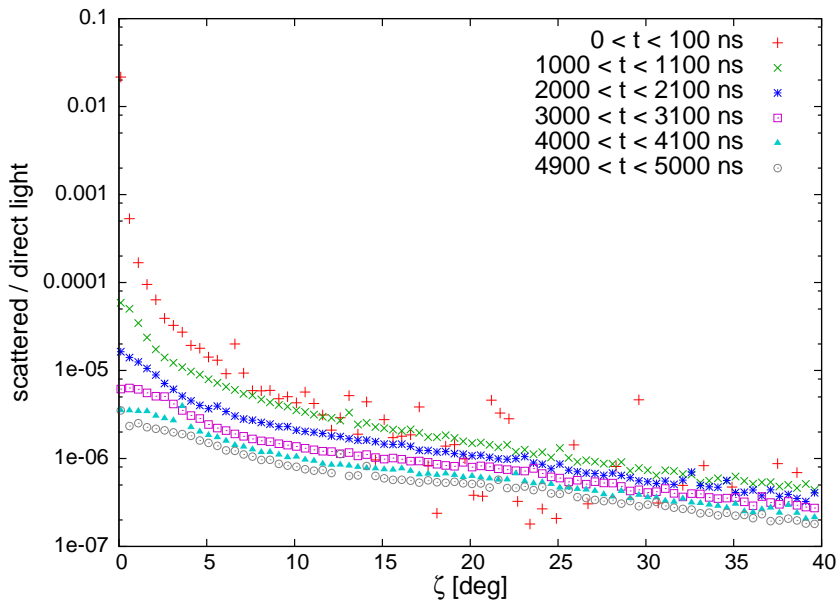}\\
\caption{\label {fig1} An example of results from one simulation. The light source is 7 km from the detector, at an altitude of 3 km. Aerosol scattering is described by the Longtin phase function. Presented are distributions of scattered light arriving to the detector as a function of angle $\zeta$ from the position of the source at different time bins. The scattered light is integrated over 100 ns time bins, and calculated per square degree.}
\end{center}
\end{figure}

An example of the scattered light distributions produced by simulations for one geometry and one atmosphere is presented in fig. \ref{fig1}. On this and other plots the ratio of scattered to direct light is presented. This ratio has no physical units, but one should keep in mind that it is calculated for 100 ns time bins and per unit square degree of solid angle. The logarithmic scale demonstrates that the intensity of light falls off steeply with the distance from the center of the image. This is seen most prominently in a few first time bins of observation. Because of the short time elapsed since the start of the emission, the light can not travel far off, so the signal is dominated by photons scattered by small angles, traveling close to the direct source-to-detector line. For later time bins, the scattered signal fades down slowly with angle, and is no longer so strongly peaked at the center of the image.

The results of the simulations allow one to quickly estimate how important is the effect of multiple scattering in different conditions. Our previous analysis \cite{pekala09} showed that the most significant contribution of scattered light is observed for large distances and small altitudes. This has been confirmed in the current simulations. For the Longtin aerosol phase function, in the central region of a radius of $1^{\circ}$, the integrated scattered signal reaches about 1\% intensity of the direct signal for the most distant sources (35 km); for smaller distances ($\simeq$ 10 km) this contribution is smaller, of the order of 0.5\%. One should remember, that in air shower observations the energy estimations are based on the measured light intensity, and accounting for the scattered light causes a corresponding shift of the energy scale. The largest contributions occur only for a short time after the start of observation, when the light is strongly peaked at the center of the image, and so increasing the opening angle does not lead to a proportional increase of the signal: within a radius of $5^{\circ}$ it may be about twice as large as within $1^{\circ}$. Only a few degrees off the center, the intensity of scattered light falls by an order of magnitude. After the light source stops emitting, the scattered light signal falls down, by about an order of magnitude within the first microsecond.

\begin{figure}[ht]
\begin{center}
\includegraphics[scale=0.9]{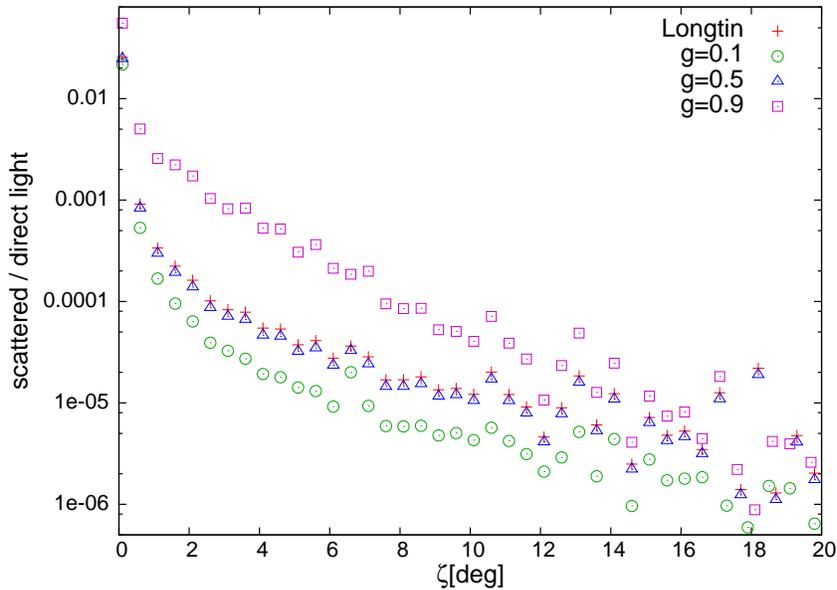}\\
\caption{\label {fig7} An example of results from simulations for different aerosol phase functions. In all simulations the light source is 7 km from the detector, at an altitude of 3 km, at the same atmospheric conditions (except different phase functions). Presented are distributions of scattered light arriving to the detector during first 100 ns of observation.}
\end{center}
\end{figure}

To see what is the effect of the aerosol phase function on the observed scattered signal, we can compare the results from respective simulations. In fig. \ref{fig7} are shown results from simulations at exactly the same geometries and aerosol concentrations, only the aerosol phase functions were different. These distributions differ in amplitude of the forward scattering peak, and so at the start of observations, when mainly scattering by small angles contributes to the signal, also the effects of different phase functions are most prominent.
The phase functions with stronger forward peak cause also larger contributions of scattered light: for the Henyey-Greenstein function with parameter $g$=0.1 the scattered light signal is the smallest (less than 1\% in the central circle of $1^{\circ}$ in all conditions), while for $g$=0.9 it is significantly larger (in extreme cases more than 10\% within radius of $1^{\circ}$).

\section{Parameterization of contribution from scattered light}

It is possible to make dedicated simulations of the scattering of light for each experimental setup, but it would require much time to investigate all conditions, at which showers are observed. In a general case of extensive calculations, in various geometries and atmospheric conditions, a parameterization is required. Such a parameterization should allow the experimenters to quickly calculate the distribution of scattered light based on a set of parameters that describe the conditions of the observation.

\begin{figure}[ht]
\begin{center}
\includegraphics[scale=0.9]{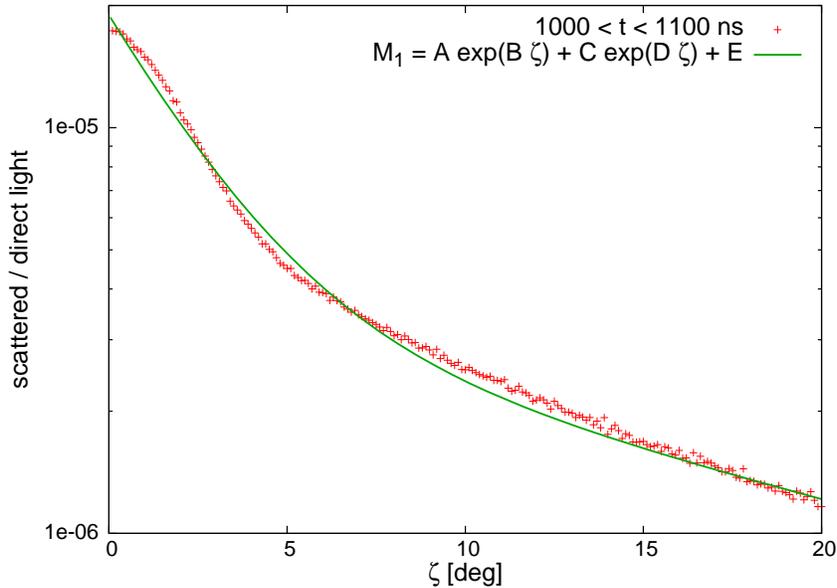}\\
\caption{\label {fig2} An example of a fit to results from one simulation for late time of observation ($t > 200$ ns). A sum of two exponentials is fitted to the simulated data for a time bin $1\mu$s after the start of observation.}
\end{center}
\end{figure}

\begin{figure}[ht]
\begin{center}
\includegraphics[scale=0.9]{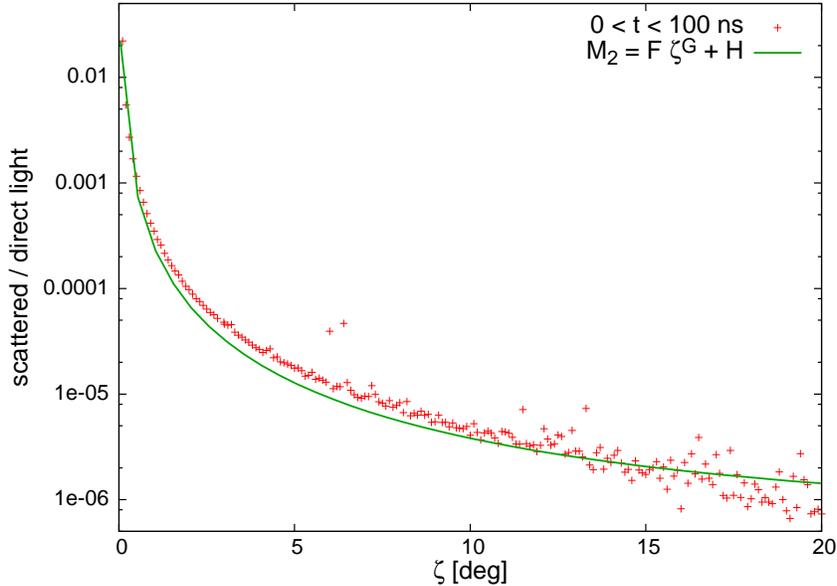}\\
\caption{\label {fig3} An example of a fit to results from one simulation for early time of observation ($t < 200$ ns). A power function is fitted to the simulated data from first 100 ns of observation.}
\end{center}
\end{figure}

The parameterization must describe the distributions of light on the sky, preferably not only in the vicinity of the location of the source, but also for more distant areas. In this study many mathematical functions of different forms were examined, to find out which could best match the scattered light distributions at all the different conditions of observation. A feature that proved to be the most difficult to describe was the change of the distribution shape in time. As it was mentioned before, at the start of the observation the scattered light is strongly collimated near the location of the source; later the distribution becomes wider, with smaller central peak. No single simple function could satisfactorily describe the shape of the distribution at all times of observation, therefore after investigating an extensive set of potentially matching formulas, two functions were chosen to parameterize the scattered light distributions. For times of observation greater than 200 ns, a sum of two exponentials of form $M_2 = D \exp(E \zeta) + F \exp(G \zeta) + H$ is used. It describes well the shape of the distribution in this longer period of time (fig \ref{fig2}). Another function, of the form $M_1 = A \zeta^B + C$ is used for the first 200 ns. It matches well the strong peak of scattered light in the initial time bins of the observation (fig \ref{fig3}). We remind the reader that  $M_1$ and $M_2$ represent the distribution of the scattered light normalized by the intensity of the direct light in the first time bin.

The $M_1$ and $M_2$ functions describe the intensity of scattered light from different directions, arriving simultaneously to a detector located on the ground. The complete parameterization must also include other parameters, that describe the conditions of observation, as well as time. Our previous analysis \cite{pekala09} had shown that the altitude above ground and the optical distance from source to detector can be used to parameterize the contribution of scattered light. Using the optical distance $\tau$ measured in units of mean free path, rather than the geometrical distance in kilometers, allows us to include uniformly simulations for different atmospheric conditions (aerosol concentration), without need for another parameter. It also enables using these results, that are obtained from simulations of a monochromatic light source, to sources of various spectra, after correcting the optical depth for different wavelengths.

The final parameterizations are functions of angle $\zeta$ from the center of the image, time $t$ from the start of observation, the source altitude above ground $h$ and the optical distance $\tau$ for the line between the source and the detector. Once the form of the parameterization $M_1$ and $M_2$ as functions of angle $\zeta$ were chosen, the dependence of their parameters ($A$ through $H$) on the other variables (time, distance, altitude) was determined. And so for example, in function $M_2$ the parameters $D$ through $H$ fall down with time, each with a specific power index. The final fits of the parameterizations were made for different aerosol phase functions separately. The formulas of the parameterization are given in the appendix.

\begin{figure}[tp]
\begin{center}
\includegraphics[scale=0.9]{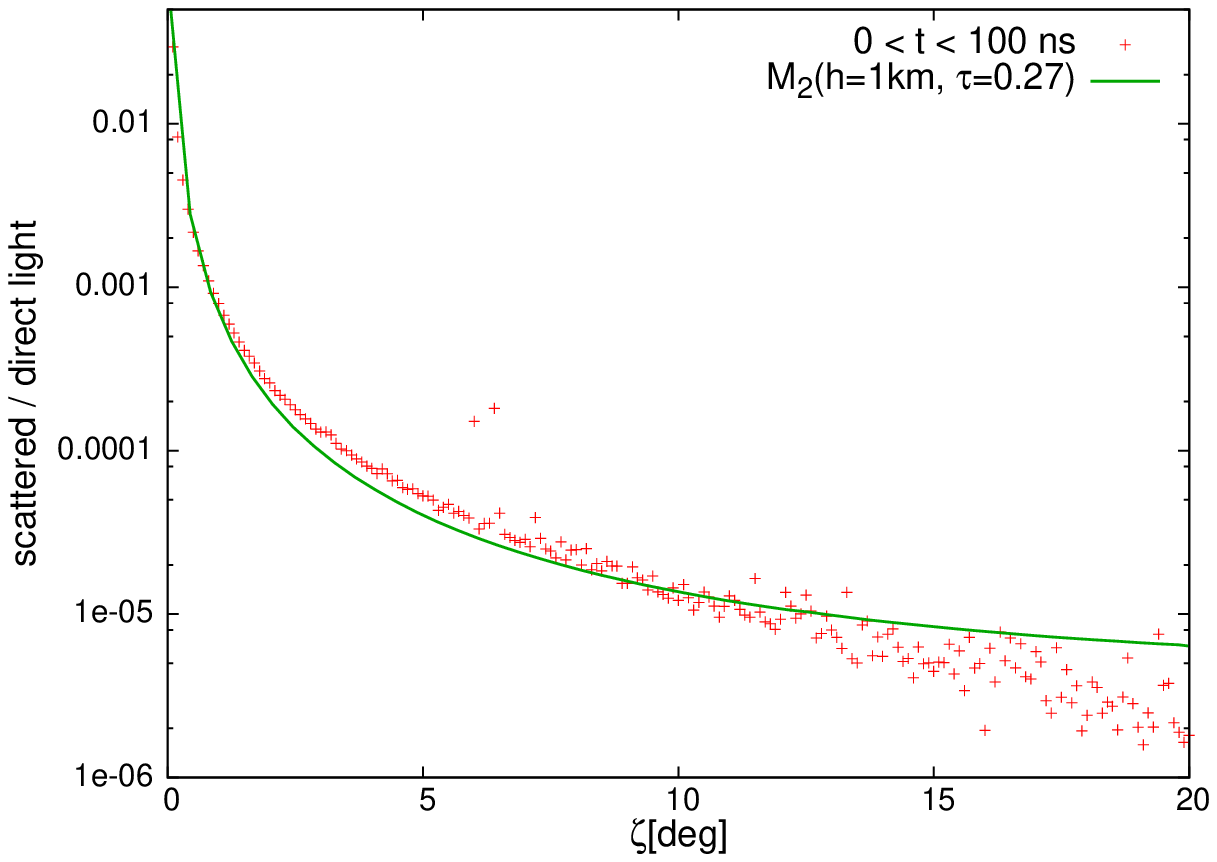}\\
\includegraphics[scale=0.9]{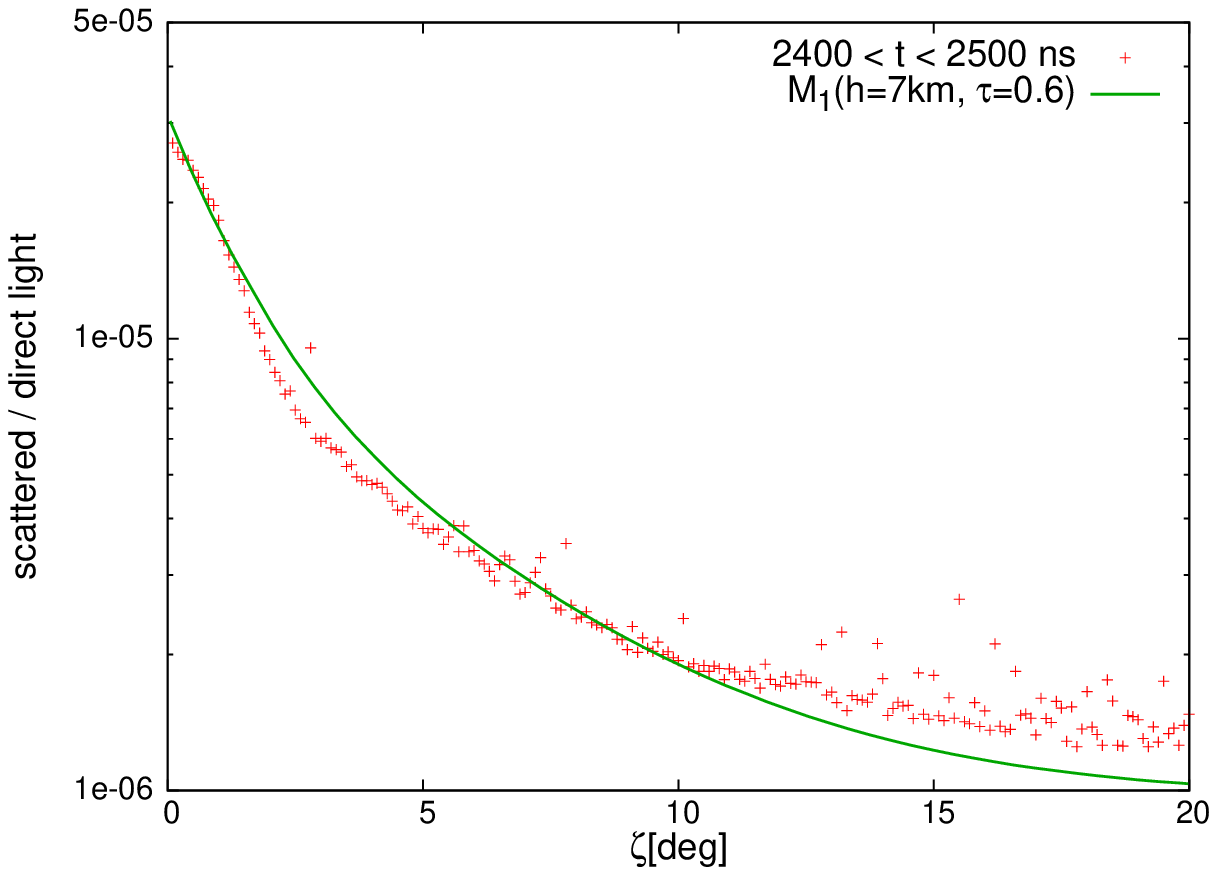}\\
\caption{\label {fig4} An example of comparison of results from the simulations using the Longtin phase function with the final parameterization.}
\end{center}
\end{figure}

This parameterization reproduces well the results of the simulations. Some examples of comparison of the final fit to the simulation results are shown in figure \ref{fig4}. The fit has been performed on data reaching out to $90^{\circ}$ off the image center, however it is most accurate out to angle $\zeta$ of about $20^{\circ}$. Beyond $20^{\circ}$ the fit may deviate from the simulated distributions by even an order of magnitude, but considering that the intensity of the scattered light at such angles ($10^{-5}$ or less of the direct light) is practically unmeasurable, this shouldn't cause any significant error in practical applications. For larger distances, corresponding to $\tau \gtrsim 0.25$, the parameterization can be used reliably: typically at the image center (within $\simeq 5^{\circ}$) the deviations are smaller than 20 \%, further out to about $20^{\circ}$, where the signal itself is much smaller, the fit does not deviate by more than a factor of 2. However, for the smallest distances ($\tau \lesssim 0.25$) the fit can deviate from the simulations also for smaller $\zeta$ angles: already at a radius of about $5^{\circ}$ the difference can be as large as 100 \%, so one should be cautious when using the parameterization.

The signal due to the scattered light calculated with this parameterization for time $t$ gives the intensity, per deg$^2$, in the period of time ($t$, $t$+100 ns). The length of a time bin in this simulation was 100 ns, and so the results describe the observations with such resolution. When applying this parameterization to studies with time bins other than 100 ns, one should be careful, especially considering the short time at the start of observation, when the changes in the distribution of light are most rapid.

\section{Conclusions}

The parameterization obtained in this work describes blurring of the image of point source of light (the point spread function) due to multiple scattering as a function of time, distance of propagation in the atmosphere, and altitude of the light source, for different aerosol phase functions (i.e. different aerosol sizes). It is normalized respectively to the direct light, therefore it is independent of the source brightness. The parameterization of the signal from multiple scattering of light in air has been developed based on an extensive set of simulations. To describe with a satisfying accuracy the distributions of light over large area of sky at different conditions and their changes in time, it was necessary to use a parameterization of a rather complicated form. Nevertheless, once the formula is implemented into a computer program, it enables easy and fast calculations of scattered light signal, from only few parameters describing the conditions of observation. It can be used in simulations of extensive air shower observations or any other observations of light sources in the atmosphere. 

The parameterization describes the distributions of light arriving to a detector on the ground, independent of any properties of the detector itself. Therefore the parameterization can be applied in modeling observations in any optical detector. It will allow experimenters to better estimate the background of scattered light in the signal. Also, one can estimate the part of the signal that is lost in the detector in the pixels below the trigger level. As it was demonstrated with the help of previous studies on multiple scattering of light, this effect causes significant systematical changes of the energy estimations of air showers \cite{benzvi10}, so using this parameterization will help decrease the uncertainty of air shower experiments using fluorescence detectors.

In this work we have investigated scattered light from an isotropic source. In case of the Cherenkov radiation the light emission is not isotropic, but highly collimated, so the geometry of scatterings is different, and it should be expected that the distributions of scattered light will also differ. Therefore the parameterization presented in this work can not be directly applied. The scattering of Cherenkov light and their impact on the atmospheric Cherenkov technique has been already investigated \cite{bernloehr00}. However, in this case the direct Cherenkov dominates the observed signal, while in air shower observations with fluorescence detectors in most cases not the direct, but scattered Cherenkov light contributes significantly. Detailed description of this effect would require a separate analysis that could be done using a modified version of our program.

\textit{Acknowledgements.} We would like to thank Ralph Engel for useful discussions and help in preparing this work. This work was partially supported by the Polish Ministry of Science and Higher Education and National Science Centre under grant No. N N202 2072 38 and by the German Academic Exchange Service (DAAD) under grant No. 507 255 95.

 \appendix
\section{Parameterization of the point spread function}

In the formulas below angle $\zeta$ is in degrees, time $t$ is in nanoseconds and altitude above ground $h$ in kilometers.

\begin{equation} \label{functions}
\begin{split}
t &< 200ns:\;
 M_1 = A \zeta^B + C \\
t &\geq 200ns:\;
 M_2 = D \exp(E \zeta) + F \exp(G \zeta) + H \\
\end{split}
\end{equation}

For the Longtin phase function:
\begin{equation} \label{fitdeft0}
\begin{split}
 A &= (0.00152 + 1.65 \cdot10^{-5} \times t) \times \tau^{0.0826 + 0.00525 t} \times \exp(-0.644 h) \\
   & + 5.69 \cdot10^{-5} + 1.1 \cdot10^{-6} \times t \\
 B &= -0.0504 h \tau + (-0.170 + 0.000766 t) \tau - 0.0323 h - 1.78 + 0.00814 t \\
 C &= 3.59 \cdot10^{-6} \times \tau + (2.03 \cdot10^{-7} + 1.65 \cdot10^{-9} \times t) h \\
   & + 2.44 \cdot10^{-6} - 6.2 \cdot10^{-8} \times t
\end{split}
\end{equation}

\begin{equation} \label{fitdef}
\begin{split}
 D &= 128 \tau \times \exp(-0.329h) \times t^{-1.76} \\
 E &= (-81.3 \tau + 21.8) h \times t^{-0.857} \\
 F &= 2004 \tau \times \exp(-0.103 \times h) \times t^{-2.22} \\
 G &= (-132 h \tau + 78.4 h - 600) \times t^{-0.833} \\
 H &= 0.0207 \times \exp(-0.367 \tau) \times \exp(-0.371 h) \times t^{0.0812 h - 1.54}
\end{split}
\end{equation}

For the Henyey-Greenstein function with $g$=0.1:
\begin{equation} \label{fit1t0}
\begin{split} 
 A &= (0.000272 + 7.88 \cdot10^{-6} \times t) \times \tau^{-0.248 + 0.0061 t} \times \exp(-0.456 h) \\
   & + 4.04 \cdot10^{-5} + 1.29 \cdot10^{-6} \times t \\
 B &= -0.0372 h \tau + (-0.272 + 0.000413 t) \tau - 0.00927 h - 2.04 + 0.009 t \\
 C &= 1.61 \cdot10^{-6} \times \tau + (-1.56 \cdot10^{-8} + 2.82 \cdot10^{-9} \times t) h \\
   & + 1.64 \cdot10^{-6} - 5.09 \cdot10^{-8} \times t
\end{split}
\end{equation}

\begin{equation} \label{fit1}
\begin{split}
 D &= 43 \tau \times \exp(-0.098 h) \times t^{-1.72} \\
 E &= (-55.4 \tau + 18) h \times t^{-0.829} \\
 F &= 2275 \tau \times \exp(-0.0586 \times h) \times t^{-2.26} \\
 G &= (-151 h \tau + 78.6 h - 457) \times t^{-0.814} \\
 H &= 22.1 \times \exp(-9.99 \tau) \times \exp(-2.99 h) \times t^{0.0258 h - 0.881}
\end{split}
\end{equation}

For the Henyey-Greenstein function with $g$=0.5:
\begin{equation} \label{fit5t0}
\begin{split}
 A &= (0.00123 + 1.6 \cdot10^{-5} \times t) \times \tau^{0.0346 + 0.00544 t} \times \exp(-0.637 h) \\
   & + 5.49 \cdot10^{-5} + 1.15 \cdot10^{-6} \times t \\
 B &= -0.0495 h \tau + (-0.186 + 0.000749 t) \tau - 0.0292 h - 1.82 + 0.00831 t \\
 C &= 2.98 \cdot10^{-6} \times \tau + (1.41 \cdot10^{-7} + 2.03 \cdot10^{-9} \times t) h \\
   & + 2.3 \cdot10^{-6} - 5.62 \cdot10^{-8} \times t
\end{split}
\end{equation}

\begin{equation} \label{fit5}
\begin{split}
 D &= 90.7 \tau \times \exp(-0.215 h) \times t^{-1.77} \\
 E &= (-74.6 \tau + 19.8) h \times t^{-0.863} \\
 F &= 4260 \tau \times \exp(-0.297 \times h) \times t^{-2.24} \\
 G &= (-130 h \tau + 77.2 h - 613) \times t^{-0.829} \\
 H &= 0.0141 \times \exp(-0.286 \tau) \times \exp(-0.415 h) \times t^{0.0748 h - 1.44}
\end{split}
\end{equation}

For the Henyey-Greenstein function with $g$=0.9:
\begin{equation} \label{fit9t0}
\begin{split}
 A &= (0.0128 - 4.4 \cdot10^{-5} \times t) \times \tau^{0.607 + 0.0026 t} \times \exp(-0.43 h) \\
   & + 0.000474 - 5.66 \cdot10^{-6} \times t \\
 B &= (0.0343 - 0.000747 t) h \tau + (-0.178 + 0.0023 t) \tau \\
   & + (-0.0132 - 0.000584 t) h - 1.33 + 0.00665 t \\
 C &= (-4.34 \cdot10^{-6} - 1.35 \cdot10^{-6} \times t) \tau + (4.73 \cdot10^{-6} + 2.61 \cdot10^{-7} \times t) h \\
   & - 1.18 \cdot10^{-5} - 8.11 \cdot10^{-7} \times t
\end{split}
\end{equation}

\begin{equation} \label{fit9}
\begin{split}
 D &= 161 \tau \times \exp(-0.068 h) \times t^{-1.45} \\
 E &= (-60.2 \tau + 38.5) h \times t^{-0.244} \\
 F &= 934 \tau \times \exp(-0.0437 \times h) \times t^{-2.16} \\
 G &= (-49.5 h \tau + 164 h - 1340) \times t^{-0.824} \\
 H &= -0.0854 \times \exp(0.263 \tau) \times exp(-0.767 h) \times t^{-1.11}
\end{split}
\end{equation}







\end{document}